\documentclass[a4paper,twoside]{article}

\usepackage{epsfig}
\usepackage{subcaption}
\usepackage{calc}
\usepackage{amssymb}
\usepackage{amstext}
\usepackage{amsmath}
\usepackage{amsthm}
\usepackage{multicol}
\usepackage{pslatex}
\usepackage{apalike}
\usepackage{SCITEPRESS}     
\usepackage{comment}

\begin{document}

\title{I-ODA, Real-World Multi-modal Longitudinal Data for Ophthalmic Applications}


\author{\authorname{Nooshin Mojab\sup{1}, Vahid Noroozi\sup{1}, Abdullah Aleem\sup{2}, Manoj P. Nallabothula\sup{2}, Joseph Baker\sup{2}, Dimitri T. Azar\sup{2}, Mark Rosenblatt\sup{2}, RV Paul Chan\sup{2}, Darvin Yi\sup{2}, Philip S. Yu\sup{1} and Joelle A. Hallak\sup{2}}
\affiliation{\sup{1}Department of Computer Science, University of Illinois at Chicago, Chicago, IL, US}
\affiliation{\sup{2}Department of Ophthalmology and Visual Sciences, University of Illinois at Chicago, Chicago, IL, US}
\email{\{ nmojab2, vnoroo2, aaleem2, mnalla2,jbaker7, dazar, mrosenbl, rvpchan, dyi9, psyu, joelle\}@uic.edu}
}


\abstract{Data from clinical real-world settings is characterized by variability in quality, machine-type, setting, and source. One of the primary goals of medical computer vision is to develop and validate artificial intelligence (AI) based algorithms on real-world data enabling clinical translations. However, despite the exponential growth in AI based applications in healthcare, specifically in ophthalmology, translations to clinical settings remain challenging. Limited access to adequate and diverse real-world data inhibits the development and validation of translatable algorithms. In this paper, we present a new multi-modal longitudinal ophthalmic imaging dataset, the Illinois Ophthalmic Database Atlas (I-ODA), with the goal of advancing state-of-the-art computer vision applications in ophthalmology, and improving upon the translatable capacity of AI based applications across different clinical settings. We present the infrastructure employed to collect, annotate, and anonymize images from multiple sources, demonstrating the complexity of real-world retrospective data and its limitations. I-ODA includes $12$ imaging modalities with a total of $3,668,649$ ophthalmic images of $33,876$ individuals from the Department of Ophthalmology and Visual Sciences at the Illinois Eye and Ear Infirmary of the University of Illinois Chicago (UIC) over the course of $12$ years.}

\onecolumn \maketitle \normalsize \setcounter{footnote}{0} \vfill

\section{\uppercase{Introduction}}
\label{sec:introduction}

\noindent The past decade has witnessed dramatic growth in the development of artificial intelligence (AI) applications in healthcare, specifically in ophthalmology \cite{schmidt2018artificial,lu2018applications,ting2019artificial,grewal2018deep}. With the promising success of deep learning models in computer vision, the field of medical imaging analysis has grown immensely towards the development of deep learning based applications serving multiple purposes. Several research studies have employed deep learning algorithms to address various problems in ophthalmology from detection to progression
predictions \cite{burlina2017automated,burlina2018use,varadarajan2018deep,gargeya2017automated,gulshan2016development,medeiros2019machine,thompson2019deep,fu2018disc}. Despite the high performance of these models, their translation to real-world clinical settings is still an ongoing problem.
Our goal is to address three core research problems in ophthalmic computer vision applications: (i) advance medical computer vision and machine learning-based applications in ophthalmology; (ii) provide an infrastructure to enhance generalizations and the translational capacity of AI applications across different clinical settings; and (iii) understand disease progression trends across various ophthalmic diseases, and
address algorithm bias. 
\begin{figure*}[ht]
    \centering
    \includegraphics[width=0.75\textwidth]{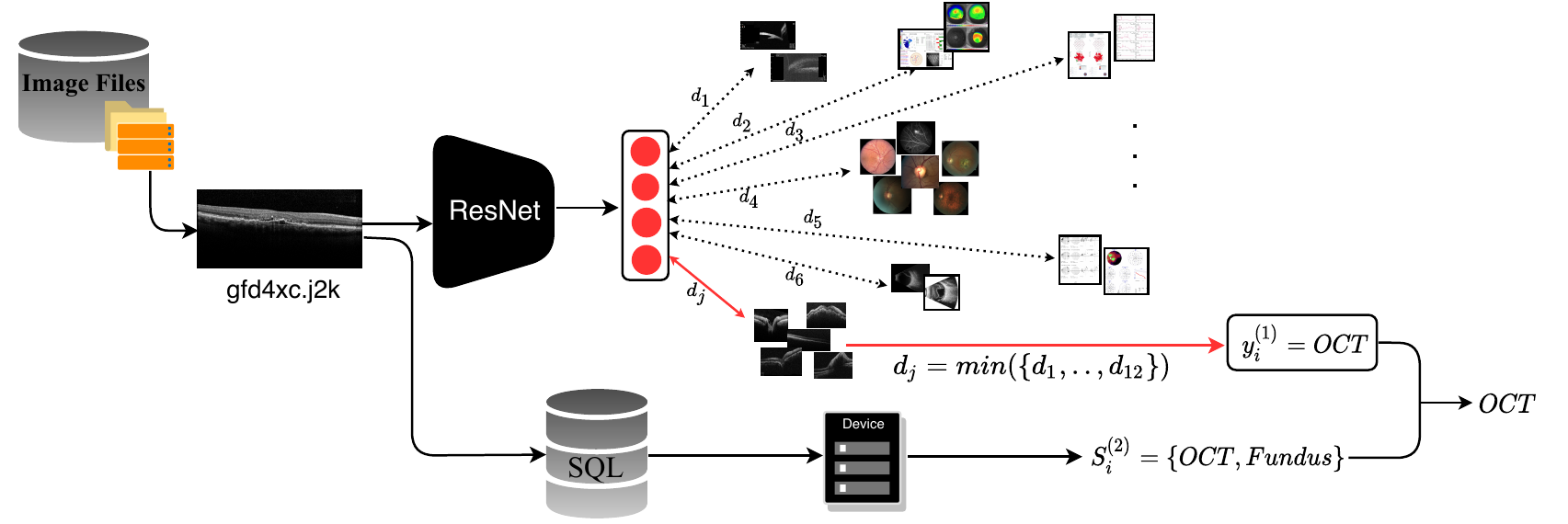}
    \caption{Illustration of the overall pipeline network for modality tagging. The raw images are fed to a ResNet network and the modality tag is achieved by comparing the input image features with prototype images, which are further validated by device information.}
    \label{fig:img_pipeline}
\end{figure*}
One of the main limitations with AI applications in healthcare is the lack of adequate and diverse longitudinal patient data to allow for successful translational applications. The current publicly available datasets that are mostly used to develop AI algorithms for ophthalmic applications \cite{fumero2011rim,almazroa2018retinal,sivaswamy2014drishti,decenciere2014feedback} have three main limitations: (i) limited number of patients and imaging data, where most of the data come from artificial settings such as multi-center clinical trials, (ii) lack of longitudinal imaging data from various modalities, and (iii) the potential risk of bias emanating from the lack of diversity in patient data. 

Building a research-oriented medical imaging databank from real-world settings could potentially improve generalizations and clinical translations. However, there are several challenges in building real-world imaging and clinical
datasets including: (i) limited access to original raw data due to Health Insurance Portability and Accountability Act (HIPAA), patient privacy, and ambiguity in data ownership, (ii) data sources across multiple heterogeneous settings with insufficient information on the data description, collection process and integration, (iii) lack of ground truth labels and standardization, and (iv) complex anonymization process with strict data sharing regulations.

In this paper, we introduce a longitudinal multi-domain and multi-modal imaging dataset for ophthalmic applications, the Illinois Ophthalmic Database Applications (I-ODA). We propose an infrastructure to collect, preprocess, annotate, and anonymize image data from multiple sources. The dataset release is pending legal approval. 

Our dataset is characterized by four main key points: (1) more than $3.5$ million image instances clustered into a diverse set of practical image modalities for ophthalmic applications, (2) longitudinal data that includes patients receiving continuous care at one academic medical center, (3) a mixture of data from multiple imaging devices representing a multi-domain data, and (4) a broad disease spectrum across multiple ophthalmic diseases. 
The unique properties of our dataset capture different characteristics of a real-world clinical setting that can serve multiple purposes for AI based ophthalmic applications. I-ODA can provide an ideal infrastructure for validation studies and translations to patient care settings enabling breakthroughs in medical computer vision.

\begin{figure*}[]
    \centering
    \includegraphics[width=0.74\textwidth]{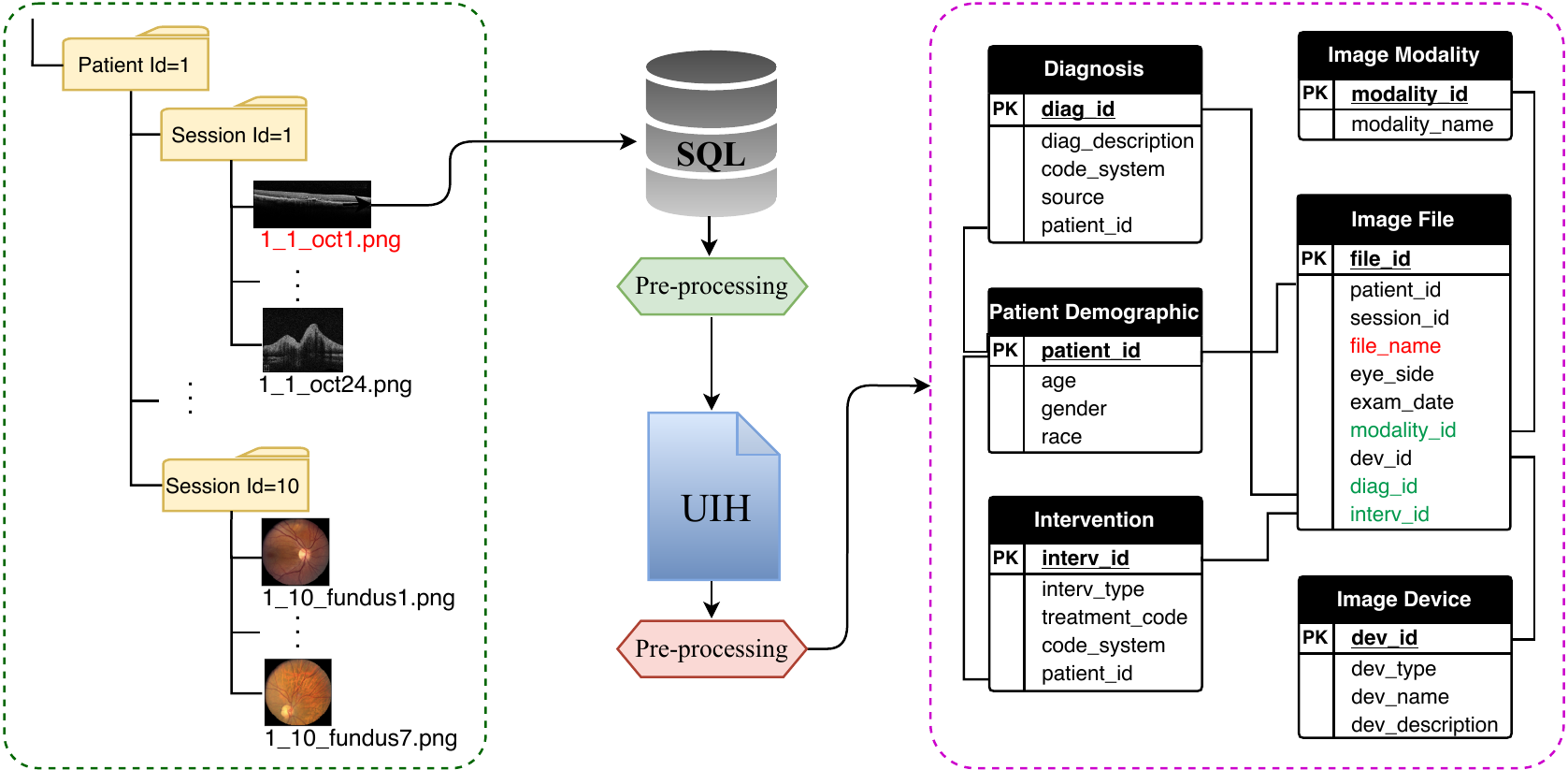}
    \caption{Illustration of the pipeline network for disease annotation and integration of the three data sources, image file hierarchy (depicted in green bounding box), SQL data that includes patient metadata, billing data that includes diagnoses and interventions. The final data schema is presented as a relational database constituting of $6$ tables (depicted in a pink bounding box) on the right side of the figure. Each image (depicted in red) is annotated with its corresponding image modality, diagnosis, and interventions (depicted in green).}
    \label{fig:diag_pipeline}
\end{figure*}

\section{\uppercase{Database and Atlas Components}}

\noindent The Institutional Review Board (IRB) of the University of Illinois at Chicago approved the creation of the I-ODA databank. Each project that utilizes the I-ODA dataset will undergo additional review by the IRB to ensure patient privacy and protocol adherence. The research to build the I-ODA dataset was conducted in accordance with the requirements of the Health Insurance Portability and Accountability Act (HIPAA) and tenets of the Declaration of Helsinki. 

The I-ODA dataset includes imaging, diagnoses, and clinical data from the Department of Ophthalmology and Visual Sciences at the Illinois Eye and Ear Infirmary of the University of Illinois at Chicago (UIC) over the course of $12$ years. The original data resides across three main sources: (i) an in-house server maintaining $\sim4.5$ million raw images belonging to $\sim45K$ patients, (ii) a SQL database consisting of patient metadata and their corresponding imaging sessions, and (iii) the University of Illinois Hospital and Health Sciences System (UIH) billing system that includes ophthalmic and non-ophthalmic diagnoses, demographics, and interventions (clinical procedures).

The raw image files are organized in a hierarchical structure sorted by Medical Record Numbers (MRNs), corresponding exam sessions, and image files residing on the in-house server that is connected to an image management system. The image files are generated by multiple imaging devices in the form of either a raw image of the eye or an analysis report. All the image files are originally stored in $.j2k (JPEG 2000)$ file format with a broad range of image resolutions.
During each visit, patients can undergo multiple imaging test sessions for each eye. Based on a preliminary diagnosis identified by an ophthalmologist, photos representing different structures can be taken from multiple angles in each imaging test session. We refer to these images as "image modalities" which can be generated from different devices. For example, a patient may require Fundus imaging, a photo of the posterior part of the eye, or Optical Coherence Tomography (OCT) imaging, which can represent high-resolution cross-sectional images of the retina. Fundus photos or OCT images are referred to as two types of imaging modalities. The modality of the image and the number of images taken per exam session could vary for each patient depending on the preliminary diagnosis. All image files are assigned with random file names that do not reveal the modality of the images they belong to, i.e. OCT or Fundus.

Our SQL Server database consists of a collection of comprehensive information including but not limited to, patient demographics and their corresponding exam sessions, images taken in each session, and the imaging device generating the image files. However, to the best of our knowledge, there is no descriptive information available on the structure of each table. Moreover, there is no constraint defined for the attributes constituting each table or the relation among the tables. This may result in invalid, missing, and duplicate data records. The billing data contains information on ophthalmic and non-ophthalmic diagnoses, procedures and interventions, and demographics. As with any billing data, it may also contain invalid or errors in data records.

Given the complexity of retrospective data, we formulated the creation of I-ODA into three main phases: (1) Modality tagging, where we tagged each image with its respective modality employing deep learning and imaging devices. (2) Disease annotation, where we annotated each image with its corresponding patient metadata, diagnosis, and intervention. (3) Anonymization, where we de-identified the whole dataset by mainly employing clustering methods, to remove any identifiable information adhering to HIPAA regulations.

\section{\uppercase{Modality Tagging}}

\noindent In this section, we describe the method used to tag all the image files with their respective modality. Due to the lack of any explicit information regarding the modality of image files, we first drafted a set of potential image modalities. Next, we yielded a set of prototype images per image modality. Lastly, given the set of modalities and prototypes, we tagged each image with its proper modality. 

\subsection{Image Modality Selection}

\noindent For the purpose of this paper, image modalities are defined as the most common imaging types used in ophthalmology. The selected set must encompass all representative modalities relevant to ophthalmic imaging applications and its diagnostic usage. 
There are more than $15$ different imaging devices in use at the Illinois Eye and Ear Infirmary at UIC, where each is responsible for generating certain image modalities. However, this assumption might be violated in a few cases. Moreover, the set of image modalities generated by each imaging device is not necessarily exclusive. For example, two different modalities, Fundus and OCT, can be generated by three different devices. As the imaging devices do not necessarily generate one modality of images, they cannot be solely used for selecting the relevant modalities but can be further utilized as auxiliary information to narrow down the potential candidates. Therefore, we first drafted a set of all potential image modalities generated by each device by extracting the imaging device information from our SQL database. Next, to keep the specificity level of each modality relevant to its diagnostic use in ophthalmology, we merged the relevant ones to enable a practical collection of image modalities with a reasonable amount of instances per modality. To achieve this goal, we selected a random subset of images per imaging device extracted from the SQL database. Given the preliminary set of modalities, we combed through the images in each subset and selected a set of relevant modalities per subset. We further reviewed the overall obtained modalities from each subset to potentially merge the relevant ones into one group. For instance, images illustrating analysis reports containing OCT and Fundus images, one image modality referred to as "OCT Report" was chosen to represent both of these images. This step was repeated multiple times to achieve the final set of the most common and practical modalities which was further reviewed by ophthalmologists. The final list contains $12$ image modalities. 

\subsection{Prototype Image Selection}
Given the image modalities obtained from the previous step, the next step was to collect a set of prototype images for each group of modality. Images belonging to each modality can vary in terms of color, shape, and resolution but they are all to be considered as various members of the same modality. For instance, all varieties of Fundus images including square or circular shaped or black and white or colored should be tagged as one image modality named Fundus. Thus, selected prototype images for each group of modality must form a representative set of the whole spectrum of images belonging to that imaging modality.

To achieve this goal, we first drafted a set of possible imaging devices that can generate each of the image modalities obtained from the previous step. We then selected a random subset of images from each of the devices for each modality group. This resulted in a preliminary collection of prototype images that were selected randomly for each of the $12$ modalities. To further refine the preliminary collection of prototypes for each group of the modalities, we employed a similarity-based classification method which will be elaborated on further in the following section. For the purpose of refinement of the prototype images, we employed the similarity-based method to tag a randomly selected subset of data from all groups of modalities by assigning the modality of their nearest neighbor from the prototype images in terms of euclidean distance. We then manually reviewed the results and analyzed the miss-classified ones according to the characteristic of the members of each modality group. If the miss-classification was due to the absence of that particular image variation in its corresponding set of image prototypes, that image variation was added to its corresponding prototype set. We repeated this step multiple times each time augmenting the set of prototypes if necessary until we reached a negligible error for each modality. This experiment resulted in a final collection of $253$ prototype images across $12$ image modalities. 

\subsection{Tagging}

The proposed tagging pipeline takes the raw image with undefined modality as input and achieves the modality tag in two sequential steps. (1) The first tag is achieved by employing a similarity-based classification method. (2) The obtained tag is verified by exploiting imaging devices. The overall pipeline network is illustrated in Fig. \ref{fig:img_pipeline}. 

\subsubsection{Similarity-based Classification}

Suppose we have a dataset with $N$ image instances and a set of $M$ modalities. Our goal was to tag each of the images in the dataset with one of the $M$ given modalities. We first employed a pretrained Convolutional Neural Network, ResNet-50, to extract the features for each image in the dataset and the set of prototype images. Suppose the dataset is denoted as $\mathcal{D}=\{x_1,...,x_N\}$ where $x_i \in R^k$ represents the feature vector and $k$ is its dimensionality. Given the $M$ image modalities, we defined the set of prototype images as $\mathcal{V}=\{v^{(p)}|p=1,..,M\}$ where $v^{(p)}=\{y^{(p)}_{1}, ..., y^{(p)}_{I_p}\}$, $y^{(p)}_{I_p}\in R^k$. $v^{(p)}$ represents the set of image prototypes for the modality group $p$ and $I_p$ denotes the number of instances in modality group $p$. We aimed to tag the images from the set $\mathcal{D}$ by assigning its nearest neighbor from the set $\mathcal{V}$ in terms of euclidean distance ${j_p}=argmin_{j_p}\|x_i-y^{(p)}_{j}\|$, $y^{(p)}_{j}\in \mathcal{V}$, $j=1,...,I_p, p=1,...,M$. To further ensure the reasonability of the obtained minimum distance for the input image, we chose a threshold for each modality group by investigating the reasonable distance range among its image members. If the minimum distance achieved by a euclidean measure matched the threshold, we assigned the tag for the input image $x_i$ by extracting the corresponding modality $p$ associated with the index $j_p$ in $\mathcal{V}$ denoted as $y^{(1)}_{i}= \mathcal{V}[I_{p-1}+{j}]$. We then applied the similarity-based method by comparing the images in $\mathcal{D}$ and the prototype images in $\mathcal{V}$ corresponding to all the $12$ image modalities and assigning its nearest neighbor from $\mathcal{V}$. The final set of modality tags achieved from this step is denoted as $\mathcal{Y}^{(1)}=\{y^{(1)}_{i}|i=1,...,N\}$.

\subsubsection{Modality Candidate Set}

To validate the modality tag achieved from the first step, we narrowed down the possible set of modality tags for each image by utilizing its corresponding imaging device. We considered three subsets of data according to their corresponding imaging devices and the range of image modalities generated by each device, (i) images associated with devices that are responsible for generating only one type of imaging modality, (ii) images associated with devices that generate a specific range of imaging modalities, usually two, and (iii) images associated with devices that their range of potential generated image modalities is not clear. 

Given these three groups of subsets, we assigned each image in each subset to its possible set of modality tags according to its corresponding imaging device extracted from the SQL database. The first group of images which constituted $\sim12\%$ of the data, were tagged with the one image modality generated by its corresponding imaging device. The second group which constituted $\sim78\%$ of the data, was assigned with a set of potential modality tags according to their corresponding device. The third group of images which constituted less than $1\%$ of the data was assigned with an unknown tag. The set of modality tags obtained from each of these three groups of images is denoted as $\mathcal{Y}^{(2)}=\{S^{(2)}_{i}|i=1,...,N\}$ where $S^{(2)}_{i}$ represents the set of potential image modalities for the input image $x_i$.   

Given the label sets $\mathcal{Y}^{(1)}$ and $\mathcal{Y}^{(2)}$, the final tag is assigned to each image if $y^{(1)}_{i} \in S^{(2)}_{i}$ for $i=1,...,N$. Otherwise, it is assigned as unknown for further manual review and investigation. 

\section{\uppercase{Disease Annotation}}

\noindent In this section, we describe our method for labeling each image with its corresponding patient metadata, diagnoses, and interventions utilizing our SQL server database and UIH billing data. 

\subsection{Metadata}

The SQL server database maintains a collection of comprehensive information on each individual's metadata. The data is stored across multiple SQL tables. To the best of our knowledge, there is no descriptive information on the contents of tables or the integration among the data in different tables. Therefore, we manually reviewed the set of attributes and the content in each table and isolated four tables for the purpose of creating our imaging dataset I-ODA, including patient, file, exam, and device. In each table, we only kept the attributes relevant to the creation of the I-ODA dataset and disregarded the rest. 

Due to manual entry from the imaging device interface and lack of defined constraints on tables, the data stored in the SQL tables are prone to noise and errors. Therefore, we first applied a sequence of preprocessing steps to filter out the invalid data records. The main preprocessing steps included  filtering out invalid MRNs, duplicate MRNs with different data records, missing data records across the relevant tables and mismatched information across SQL data records and image file hierarchy. The patient and file table originally contained $44,460$ patients and $4,477,634$ image files, respectively. Applying the preproccessing steps resulted in removing $\sim8\%$ of the data. Further, we integrated the data from these four tables into one file using their common attributes. 

\begin{figure*}[h]
\begin{subfigure}{.18\textwidth}
  \centering
  \includegraphics[width=.80\linewidth]{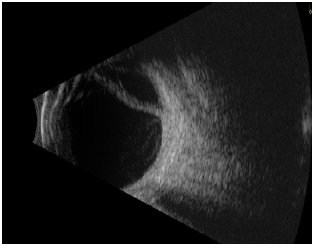}  
  \caption{B-Scan}
  \label{fig:bscan}
\end{subfigure}
\begin{subfigure}{.50\textwidth}
  \centering
  \includegraphics[width=.80\linewidth]{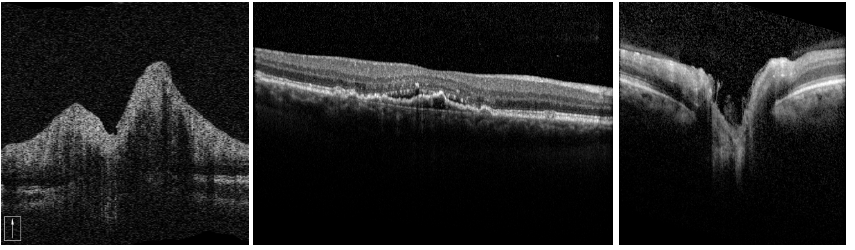}  
  \caption{OCT}
  \label{fig:oct}
\end{subfigure}
\begin{subfigure}{.24\textwidth}
  \centering
  \includegraphics[width=1.17\linewidth]{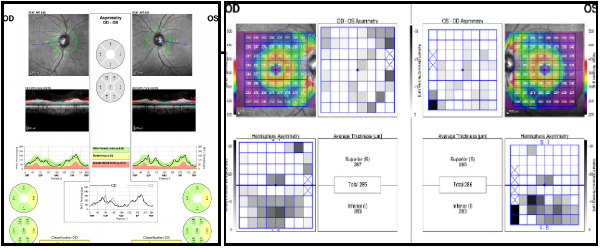}  
  \caption{OCT Report}
  \label{fig:octreport}
\end{subfigure}
\newline
\begin{subfigure}{.18\textwidth}
  \centering
  \includegraphics[width=.80\linewidth]{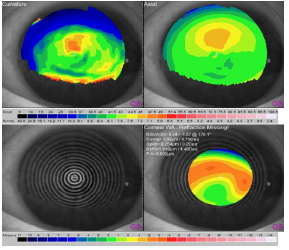}  
  \caption{Corneal Topography}
  \label{fig:ultrasound}
\end{subfigure}
\begin{subfigure}{.50\textwidth}
  \centering
  \includegraphics[width=.80\linewidth]{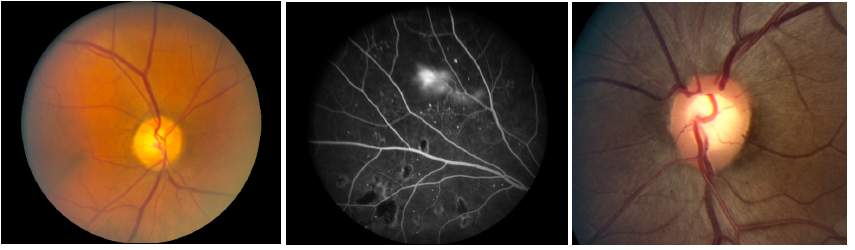}  
  \caption{Fundus}
  \label{fig:fundus}
\end{subfigure}
\begin{subfigure}{.24\textwidth}
  \centering
  \includegraphics[width=1.17\linewidth]{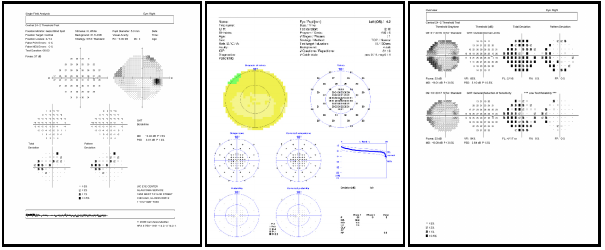}  
  \caption{HVF}
  \label{fig:hvf}
\end{subfigure}
\caption{A snapshot of samples from $6$ major image modalities in I-ODA dataset.}
\label{fig:img_categories}
\end{figure*}
\subsection{Diagnosis and Intervention}

The University of Illinois Hospital and Health Sciences System (UIH) billing system contains information regarding diagnosis and interventions. The system is equipped with a billing report and a dashboard interface allowing to retrieve hospital charges given the patient MRN for a specific date range. Given the valid patient MRNs obtained from the previous section, we extracted the corresponding charges from the billing reports to retrieve all ophthalmic and non-ophthalmic diagnoses and interventions for each patient. The interventions are referred to as any surgical or invasive outpatient or hospital procedure. 
  
First, we matched the format of MRNs in the billing system to the format of the data used in the SQL database. Next, similar to the previous section, we applied series of preprocessing steps that resulted in removing $\sim12\%$ of the data. Then we integrated the billing data with the metadata file obtained from the SQL database in the previous section. This file was further validated for any invalid or mismatched data records which resulted in excluding another $\sim6\%$ of the data. The final file contained $33,876$ patient and $3,668,649$ image files which were annotated with their corresponding metadata, diagnoses, and interventions. 

At last, we constructed a relational database integrating the data from all the data sources, image files, metadata, and diagnoses. The tables are connected through a primary key (PK) and a foreign key constraints defined for each table. Each table consists of a set of relevant attributes demonstrating its associated metadata information. The schematic of the data schema representing the relational database and data integration is illustrated in Fig.~\ref{fig:diag_pipeline}.

\section{\uppercase{Anonymization}}
\noindent Data anonymization is the process where patient identifiers are irreversibly removed for patient privacy protection, prohibiting any direct or indirect identification. According to the HIPAA regulations, sensitive patient information should be protected by being properly anonymized before being used for any research purposes. Data anonymization in the context of our work would result in a complete anonymized dataset across both data components, image files, and the associated metadata. 

\subsection{Image Anonymization}

The image members in each of the $12$ modality group in our dataset can vary in terms of style, resolution, and location of identifiable information that needs to be masked out. The extensive range of variability among images poses a major challenge on anonymization for such a large amount of data. To address this challenge, we employed a K-means clustering method to derive a set of categories for each of the $12$ modalities where the images in each category are the most similar ones in terms of style, resolution, and location of identifiable information. To choose the initial number of clusters for each modality group, we first randomly selected a subsample of $200$ images from each modality and manually reviewed and analyzed the selected subsamples. We further applied a set of various imaging filters, including spatial/geometric, resolution, appearance, and color, to achieve a more fine-grained categorization for each of the categories obtained from the initial clustering. The set of filters were chosen to be relevant to the type of images belonging to each modality. 

Next, we divided the obtained categories into two groups based on the consistency level of the location of identifiable information in each category. For the first group of categories having consistent patterns in terms of location of identifiable information across their image members, we generated a location-based masking filter specific to each category to mask out the part of the image that contained the identifiable information. For the second group of categories where the location of identifiable information varied across images, we combed through the data and manually removed the sensitive information. Eventually, the data was reviewed by two people to anonymize any missed data to ensure complete anonymization. The fine-grained categories were created merely for the purpose of anonymization. After accomplishing the anonymization process for all the image files, the categories were disregarded and only the $12$ major modalities were kept.

\subsection{Metadata Anonymization}
To de-identify the metadata, first, we extracted the set of sensitive attributes including the patient MRN, first and last name, date of birth, and exam session date. The patient first and last names were removed, and the MRN was replaced by a randomly generated number. To keep the longitudinal nature of the date of birth and exam session dates attributes, the patient's date of birth was replaced by patient's age and the date of exam session was replaced by subtracting the date of birth from the date of exam session. To integrate the anonymized metadata with the anonymized image files, the patient and exam session directories in the hierarchical structure of image files were renamed to the anonymized patient ids and exam session ids respectively.

\section{\uppercase{Dataset}}

\subsection{Data Statistics}
As of now, the I-ODA dataset \footnote{For questions related to the I-ODA dataset and for any collaboration interests please contact the author, Joelle Hallak.} contains $3,668,649$ images and $230,923$ exam sessions across $12$ image modalities of $33,876$ individuals from the Department of Ophthalmology and Visual Sciences at the Illinois Eye and Ear Infirmary of UIC for eye care. The set of image modalities includes Optical Coherence Tomography (OCT), OCT Report, Fundus, Humphrey visual field (HVF), Ultrasound, Ultrasound Report, B-Scans, Corneal Topography, External image (slit lamp), Intraocular Lens master calculation report (IOL), Optical Response Analyzer report, and ERG report. The I-ODA dataset is composed of two main data components integrated effectively to represent a structured ophthalmic imaging dataset, as shown in Fig.~\ref{fig:diag_pipeline}: (1) Anonymized image files that are tagged with their corresponding modality and are converted to $.png$ format and stored in a hierarchical structure. The highest level in a hierarchy represents a patient directory followed by its corresponding exam session and finally the imaging files that reside on the lowest level of the hierarchy. The patient and exam session directories correspond to the anonymized patient ids, and session ids from the metadata and image file names are named as "patientId\underline{\hspace{.07in}}sessionId\underline{\hspace{.07in}}modality" format. (2) A relational database that constitutes of $6$ tables representing patient demographics, image files, diagnoses, interventions, imaging devices, and image modalities integrated through primary and foreign key constraints. As Fig.~\ref{fig:diag_pipeline} suggests, the corresponding patient metadata, diagnosis, and intervention (depicted in green in the "Image File" table) for each image file (depicted in red in image file hierarchy on the left) can be easily retrieved from the tables in our relational database. 
\begin{figure}[ht]
    \centering
    \includegraphics[width=0.30\textwidth]{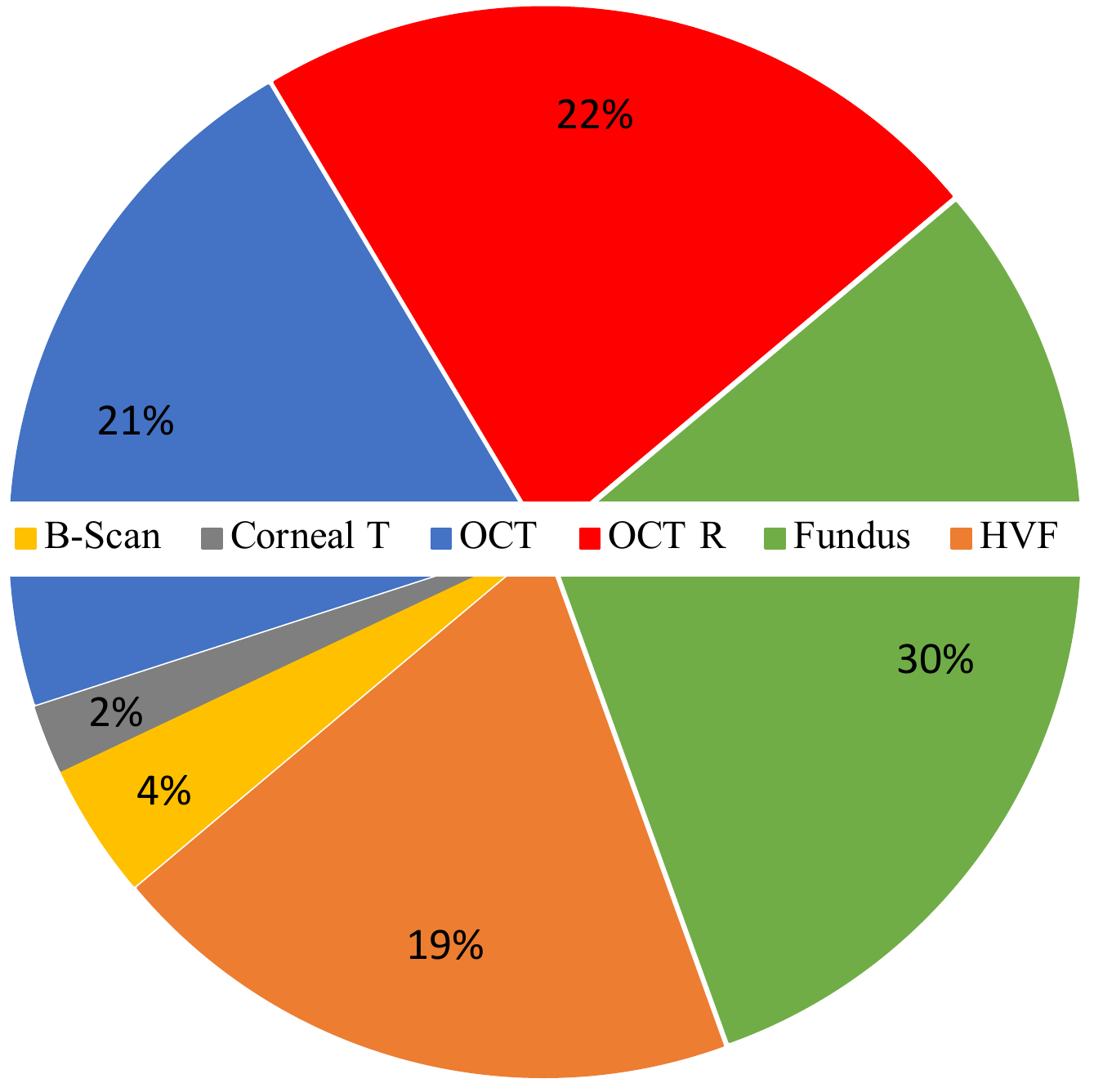}
    \caption{A snapshot of I-ODA dataset illustrating the number of imaging exam sessions for the $6$ major modalities Fundus, OCT Report (OCT R), OCT, HVF, B-Scan, Corneal Topography (Corneal T).}
    \label{fig:oda_stat}
\end{figure}
\begin{figure}[ht]
    \centering
    \includegraphics[width=0.38\textwidth]{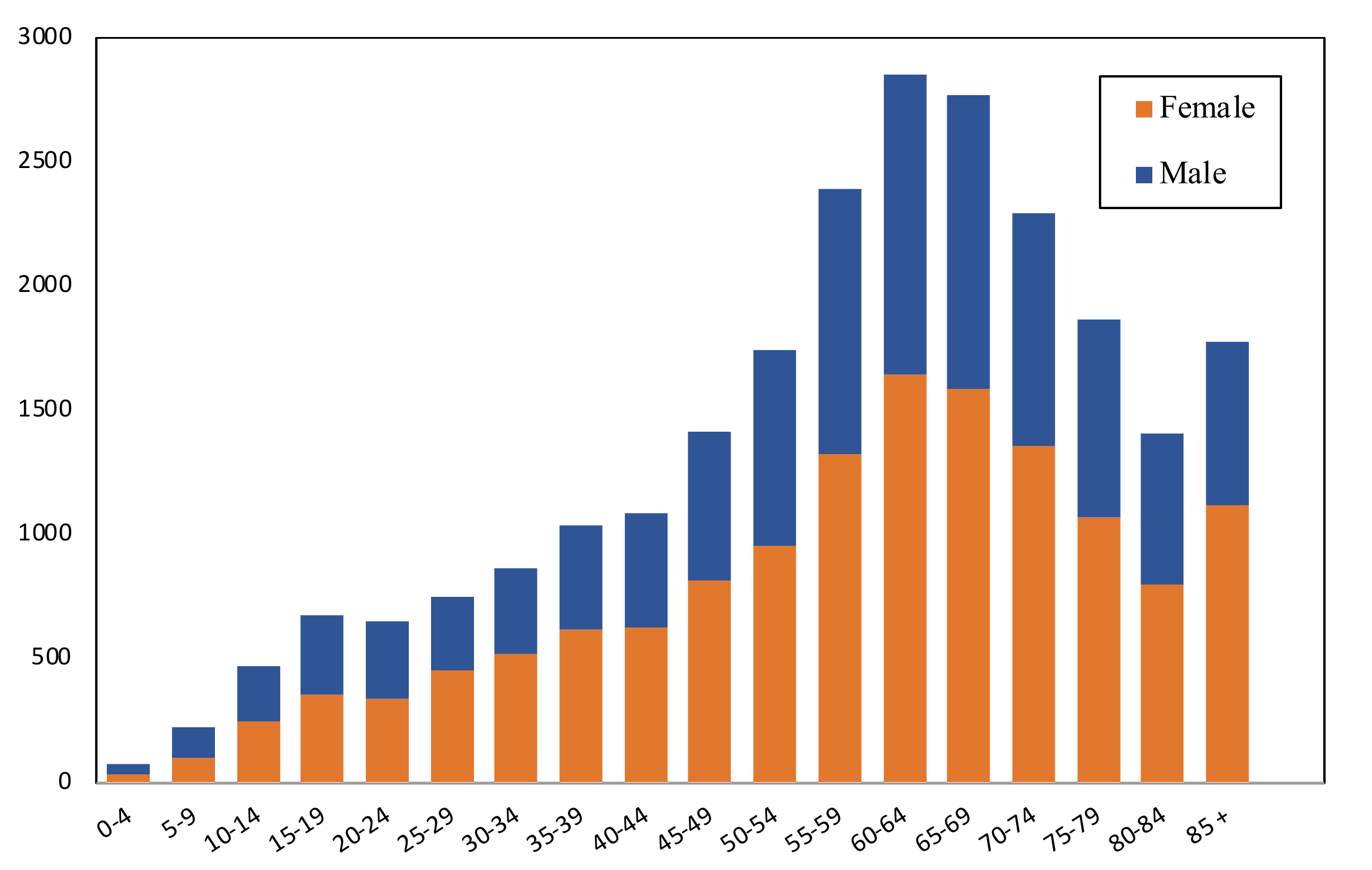}
    \caption{Illustration of the gender/age population distribution in I-ODA.}
    \label{fig:age_stat}
\end{figure}
\subsection{Data Characteristics}
Our dataset captures different characteristics of a real-world clinical setting allowing for versatile computer vision applications in ophthalmology.  
\noindent {\bf Modality and Domain:}
The I-ODA dataset comprises $12$ different modalities representing a comprehensive collection of practical image modalities relevant to ophthalmic imaging applications. Among the $12$ image modalities, $6$ modalities, Fundus, OCT Report, OCT, HVF, B-Scan, Corneal Topography, constitute $98\%$ of the imaging exam sessions. A snapshot of samples from these $6$ modalities is illustrated in Fig.~\ref{fig:img_categories}. As can be seen from the Fig.~\ref{fig:img_categories}, each image modality encompasses a spectrum of different varieties of its image members.

Ophthalmic disease imaging can include multiple sessions with different modalities per patient visit. This would result in a rich collection of longitudinal imaging sessions across different image modalities for ophthalmic applications. A summary of the I-ODA dataset showing the $6$ major image modalities and the number of exam sessions per modality is illustrated in Fig.~\ref{fig:oda_stat}. 

\begin{figure*}[ht]
    \centering
    \includegraphics[width=0.88\textwidth]{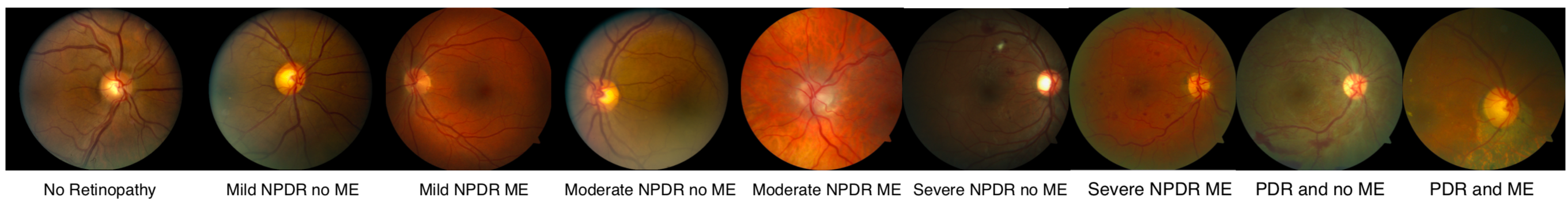}
    \caption{Illustration of the disease spectrum of DR with Fundus photos taken at different stages. NPDR represents Non-proliferative Diabetic Retinopathy, ME represents Macular Edema, and PDR represents Proliferative Diabetic Retinopathy.}
    \label{fig:dr_spectrum}
\end{figure*}
As can be seen from Fig.~\ref{fig:oda_stat}, Fundus, OCT/OCT reports, and HVF are the most commonly used imaging modalities in our I-ODA dataset. These modalities are also among the most commonly used in imaging tests for ophthalmic blinding conditions such as glaucoma, Age-related macular degeneration (AMD), and diabetic retinopathy (DR). The availability of a vast number of imaging sessions across these imaging modalities, allows us to study the disease pattern from multiple sources of data leveraging the complementary information across different views. 

Moreover, the data in I-ODA is composed of more than $15$ imaging devices forming a multi-domain dataset, where domain in here is defined as the imaging device. This important property reflects the true nature of a real-world dataset, which includes a mixture of data distributions from different domains. Given that clinical care often involves complex multi-domain data, I-ODA could provide a benchmark for validation studies and improve generalizations for translations of AI-based models across different clinical settings. 

\noindent {\bf Patient Population:}
I-ODA contains a rich collection of imaging data and metadata from a diverse set of patients with various demographic backgrounds including, ethnicity, race, age distribution, and location. A cross-tabulation analysis of patient gender and age from I-ODA is depicted in Fig.~\ref{fig:age_stat}.

As Fig.~\ref{fig:age_stat} suggests, the patient population is characterized by a comparable distribution between females and males across a wide age range. This will allow the development and validation of AI based algorithms that are a true representation of the patient population. 

\noindent {\bf Longitudinal Disease Spectrum:}
The lack of disease severity levels that do not represent the wide spectrum of patient diagnoses in real-world clinical settings can carry a risk of spectrum bias. I-ODA contains a comprehensive collection of imaging data belonging to patient visits from one academic institution across multiple time points for various ophthalmic diseases. Fig.~\ref{fig:dr_spectrum} represents the severity spectrum of Fundus photos taken at different stages of one of the ophthalmic diseases, diabetic retinopathy (DR).  

The longitudinal disease spectrum not only mitigates the risk of spectrum bias but also allows us to study the progression trends across different ophthalmic diseases. Additionally, access to imaging data for a broad range of ophthalmic diseases improves our understanding of different diseases in correlation with each other along with studying each disease in isolation. Moreover, having both ophthalmic and non-ophthalmic diagnoses in the I-ODA dataset, allows us to study potential correlations among these diseases, to identify common biological and epidemiological mechanisms.

\subsection{Application}

To demonstrate the applicability and accuracy of our dataset, we exploited different characteristics of I-ODA to address various problems in ophthalmic imaging application. 

First, we employed I-ODA to formulate the problem of glaucoma detection into a multi-task framework composed of prediction and segmentation modules with the goal of achieving interpretability and alleviating shortage of segmented data \cite{mojab2019deep}. We showed that our proposed method outperforms the strongest baseline on cup segmentation task by $2.6\%$ by utilizing the availability of adequate data from I-ODA for the prediction task. 

In our second work, we utilized I-ODA dataset to demonstrate the importance of real-world data for generalizations to clinical settings \cite{mojab2020real}. We formulated our problem into transfer learning framework employing self-supervised learning for learning visual representations. We showed the result of our work for the task of glaucoma detection by training the model on real-world data and evaluate it on a standardized data and vice versa. Our experiment showed that without training with complex multi-domain real-world data, the deep learning models do not generalize well to clinical settings. We also showed that by training our proposed method on real-world data (I-ODA), we can achieve $16\%$ relative improvement on a standardized dataset over supervised baselines. 

\section{\uppercase{Discussion}}

\noindent In this paper, we introduced a new ophthalmic imaging dataset for AI applications in ophthalmology with an infrastructure for collection, annotation, and anonymization of the data. The proposed dataset contains a diverse collection of image modalities belonging to patients who received continuous care at the Department of Ophthalmology and Visual Sciences at the Illinois Eye and Ear Infirmary at UIC. I-ODA is a longitudinal healthcare dataset that includes a large variety of ophthalmic modalities, domains, and patients. These unique properties provide an ideal infrastructure for: (i) advancements in machine learning algorithms for multi-view and multi-domain ophthalmic applications, (ii) improvements in generalizability and translations into clinical settings, and (iii) enhanced understanding of variations in ophthalmic disease prognosis. 

As a research databank with a unique infrastructure, I-ODA will continue to grow in imaging and patient metadata. While the limitations in annotations are understandable, machine learning applications developed on data from I-ODA will allow new discoveries in computer vision, specifically in the medical imaging field, and in new applications for classification and progression of ophthalmic diseases. Additionally, I-ODA can also serve with multiple efforts in validating current algorithms that have shown promise in more controlled datasets with less diverse domains and patient population.

\section{Acknowledgement }
This work is supported in part by NSF under grants III-1763325, III-1909323, SaTC-1930941, and BrightFocus Foundation Grant M2019155.

\bibliographystyle{apalike}
{\small
\bibliography{example}}

\end{document}